\documentclass[conference]{IEEEtran}
\IEEEoverridecommandlockouts
\usepackage{cite}
\usepackage{amsmath,amssymb,amsfonts}
\usepackage{algorithmic}
\usepackage{graphicx}
\usepackage{textcomp}
\usepackage{balance}
\usepackage{subcaption} 
\graphicspath{{Temperature_analysis_of_PUF_Images1/}}
\usepackage{xcolor}
\def\BibTeX{{\rm B\kern-.05em{\sc i\kern-.025em b}\kern-.08em
		T\kern-.1667em\lower.7ex\hbox{E}\kern-.125emX}}

\begin{document}
\title{A Deep Analysis of Hybrid-Multikey-PUF}

\author{\IEEEauthorblockN{ Md Ishtyaq Mahmud, Ahmed Abdelgawad,	Venkata P. Yanambaka}
\IEEEauthorblockA{College of Science and Engineering\\ 
Central Michigan University\\
}
Email: mahmu4m@cmich.edu
}
\maketitle

\begin{abstract} Unique key generation is essential for encryption purposes between Internet of Things (IoT) devices. To produce a unique key for this encryption, Physical Unclonable Functions (PUFs) might be employed. Also, the Random Number Generator (RNG) is used in many different domains; nonetheless, security is one of the most important areas that require the best RNG. In this article,  We investigate the quality of random numbers generated by Physical Unclonable Functions (PUFs). We have analyzed three Figures of Merit (FoMs), Uniqueness, Randomness, and Reliability of PUFs implemented on different FPGAs. In our experiments, we have operated the test devices at different temperatures (20\textdegree F, 40\textdegree F, 60\textdegree F, 80\textdegree F, 120\textdegree F, 140\textdegree F). In the PUF that we have analyzed, the key is generated in 1 second on average. We also have analyzed and described the essential properties of random number generator that is most vital considering things to secure our Internet of Things(IoT) devices.
  		 
\end{abstract}

\IEEEpeerreviewmaketitle


 \section{Introduction}
 In today's era of instant communication, security is crucial. Numerous cryptographic algorithms have been thoroughly evaluated and used advantageously. Many aspects of random number generation include computer programming, modeling, quantitative simulation, decision making, sampling, cryptography, etc. Most of the time, the broad idea underlying this generic term refers to sequencing, patterns, or homogeneous outputs generated by a particular source of unpredictability. Research is extensively conducted in the area of ultra-low power and low-cost Random Number Generators(RNGs) as technology advances and leads to computer machines.
 John von Neumann, who proposed a generation method based on computer arithmetic operations, was the pioneer in this field. Neumann produced numbers by subtracting the intermediate digits from the square of the previous value and then repeating the process. Mid-square is a cyclical approach that ends in a brief cycle. As a result, the key differences between the earlier generators are regularity and predictable outputs that use an arithmetician operation. "Pseudorandom" or "quasirandom" number generators (PRNGs ) are referred to in literature, whereas True" random number generators  (TRNGs) utilize a physical source to generate unpredictability TRNGs \cite{bakiri2018survey}.
 
 Random numbers are difficult to forecast, and each number created must have an equal chance of becoming an excellent contender\cite{Sathya2021}. In today's fast-paced environment, sensitive information must be maintained on the fly.  There are two ways to produce random numbers. One uses hardware modules to generate the random numbers, which use the unpredictable irregularities introduced during the manufacturing of the hardware modules to generate the random numbers. The other type is an algorithm using a seed value to generate the outputs, called Pseudo-Random Number Generators (PRNG) \cite{adewopo2022review}.
 
 
 TRNG is particularly important in the security context since it secures connections by preventing the adversary from guessing the produced numbers. TRNG generates secure, high-unpredictability true random sequences \cite{9595288}. Light, pictures, sounds, visuals, energies, and current flow, among other things, have been used to generate highly unexpected real random numbers\cite{Rando2021}. The quantum random number generator (QRNG) is a new technology that is currently being developed. Jacak et al. suggested QRNG \cite{jacak2021quantum}. QRNG necessitates the use of a separate device to produce random numbers. Furthermore, QRNG is faster than RNG generated by a machine. However, the procedure is quite difficult to complete. A PRNG is a predictable algorithm that generates a random number sequence from an initial(seed) value \cite{kietzmann2021guideline}. Because its reliance on the prior state of the process, the generated sequence is not completely random. Before being used, PRNGs must be theoretically analyzed to confirm their unpredictability qualities. This \cite{Sathya2021} article describes various PRNGs, such as Chaotic map-based PRNG, Polynomial equation-based PRANG, Hardware based PRANG, Nature Inspired PRANG, and Cryptographic algorithm-based PRANG.
 
 This paper analyzes the architecture of Hybrid-Multikey-PUF \cite{PIM2021}, a hardware security primitive used to generate one-time passcodes for resource-constrained devices. The architecture of the module was proposed in \cite{PIM2021}, but an extensive analysis was not provided in the article.
 
 The rest of the paper is presented as follows: Section II highlights the Hardware-Assisted Security (HAS) through PUF. Analysis of the PUFs architecture is presented in section III. Experimental setup and results analysis of multikey-PUFs under various temperatures is demonstrated in section IV, and section V summarizes the conclusion and future works.
  \begin{figure*}[htbp]
  	\centering
  	\includegraphics[width=1\textwidth]{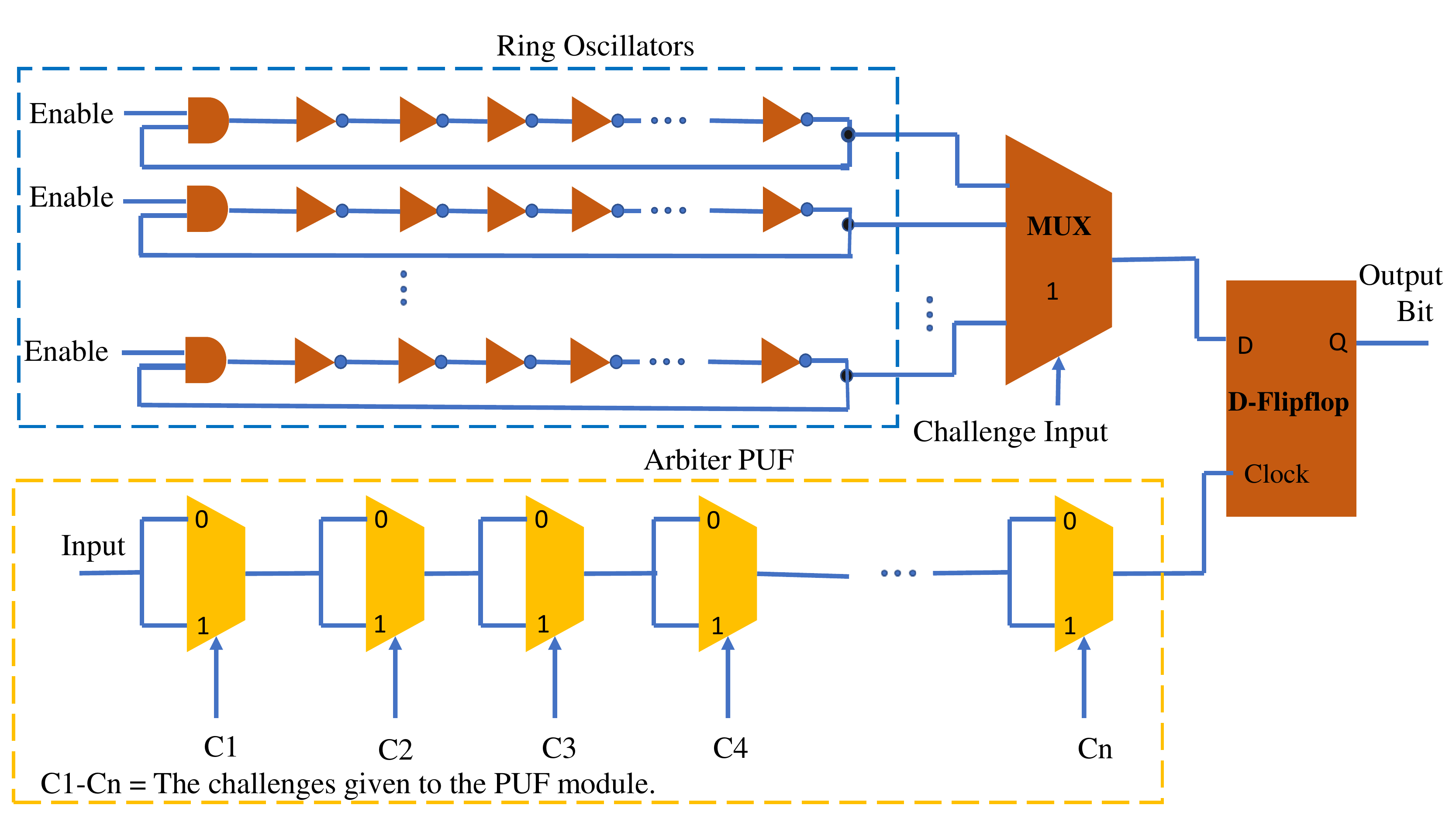}
  	\caption{Architecture of Hybrid-Multikey-PUF}
  	\label{Fig.1.}
  \end{figure*}
\section{Hardware Assisted Security through PUF}
The emerging IoT systems need robust security for user information safety. Hardware-assisted security(HAS) is one of the most vigorous security systems to secure our privacy when we are using connected devices. Numerous risks and assaults have arisen as a result of security flaws in the IoT domain, which can jeopardize vital infrastructures and public safety, cause physical and economic hardship, and more\cite{amsaad2021enhancing}. Hardware-based approaches take advantage of physical devices and can gather data to assess current software-level vulnerabilities and risk\cite{rahman2017hardware}. HAS offers hardware-based security for (1) the data to be processed, (2)  the entire system, and (3) the hardware itself \cite{mohanty2021introduction}. Yanambaka et al. introduced two PUF designs that may produce a new key each every time the circuit is performed, hence the name "Multi-Key Generator PUF" (MKG-PUF) \cite{Yanambaka2016}. They also presented concepts that were both power and speed optimum.

A random number generator (RNG) is used in cryptographic applications to encrypt and decrypt data for reliable transmission and receiving. Kalanadhabhatta et al. proposed that the starting seed is safeguarded by producing it from PUF, and physical unclonable function-based CPRNG (PUF-CPRNG) incorporates constant refreshing algorithms to guarantee that the random numbers generated are inclusion. CPRNG response signals are supplied from PUF to advanced security the PUF-CPRNG \cite{chaoticRNG2020}.
Modeling assaults have been discovered to be sensitive to these PUFs to varying degrees. Numerous scientific initiatives have tested the security of various PUF designs against modeling assaults. Idriss et al. identified the most appropriate design components in the installed PUFs. The IoT PUF security systems were then categorized, and the design of the project for each was a preceding statement on effectiveness \cite{idris2021}.

Yuan et al. suggested using blockchain to implement a PUF-based lightweight broadcasting authentication mechanism for multi-server networks. They also recommended their system, which is based on PUFs, can effectively withstand physical attacks while also providing efficient authentication and key agreement(AKA)\cite{yuang2021}. Moreover Mall et al. also suggested  AKA protocols of PUF and investigated and evaluated the benefits and drawbacks of AKA applications in the speedy domains of connected devices, WSNs, and Microgrids comprehensively and scientifically \cite{mall2022}.

  \section{Hybrid-Multikey-PUF}
  PUFs are an exciting future primitive that can be used for identification and cryptographic key storage even without the requirement of expensive devices such as secure EEPROMs\cite{6823677}. In this section, we will discuss which PUF we have analyzed under various temperatures.
  
  A PUF is based on the premise that regardless of the fact all ICs have the same mask and manufacturing process, each one is somewhat different due to normal manufacturing variability. PUFs reap the benefits of this variation to extract "hidden" information specific to the microchip (a silicon "biometric") \cite{9595111}. 
  For an identical design for each device, the delay it added is not the same. Even if each device has the same design, the delay it adds is not the same \cite{puf}. This allows the PUF to generate various keys that are exclusive to each module."challenge" is the input to the PUF, and the "response" is the output of the PUF; together, they form a "challenge-response pair" (CRP) \cite{PIM2021}.

 
 
 PUF's design is seen in Figure 1. The PUF architecture includes a Ring Oscillator (RO) PUF and a sequence of multiplexers, as shown in Figure 1. The oscillation frequencies of the oscillators have been used to produce the keys via ring oscillator PUF. The keys generated are enhanced more randomly via sending a clock pulse from the arbiter PUF to the flip-flop. This PUF is designed to create a unique key each time it is executed. In contrast to the RO PUF, an arbiter PUF is more dependable and capable of producing the same key each time it is run. When no errors must be generated in the output, a RO PUF is less dependable than an SRAM PUF or an Arbiter PUF. As a result, a RO PUF is recommended as the architecture. The oscillators and multiplexers are the most essential components of the design. 
 
 A "1" is assigned to the very first multiplexer on both inputs of the series of multiplexers. The challenge input determines which signal path to follow. The multiplexers' clock signal ensures that no pseudo-randomness is incorporated into the design. The serial multiplexers are responsible for obtaining a truly random number by adding a delay due to microelectronics manufacturing variances. The vibrations of a ring oscillator constructed on an IC are subject to numerous different variations. When the oscillations are read at the flip input, flop's errors are introduced. The configuration of the devices varies due to manufacturing differences, and no multiple inverters in the architecture are alike. Despite having the same architecture, the oscillators output distinct frequencies. Even with the same challenge, each time the program is run, a random bit is created when the clock signal from the sequence of multiplexers reaches the flip-flop.

 \section{Experimental Analysis}
 
Experimental setup for the evaluation of the PUF modules consists of designing them on Field Programmable Gate Array (FPGA) modules. A total of 5 PUF modules were designed on the FPGS for experiments. Each PUF module was subjected to the tests and the Figures of Merit were evaluated. Two different FPGAs were used in the current experimental evaluation. Arduino MKR Vidor 4000 is a microcontroller board with a Cyclone V FPGA integrated onto the board. Two PUF instances were designed on the FPGA. The compact design of the board makes it more suitable for resource constrained devices such as IoT and IoMT. 
\begin{figure*}
	\centering
	\begin{subfigure}[b]{0.3\textwidth}
		\includegraphics[width=\textwidth]{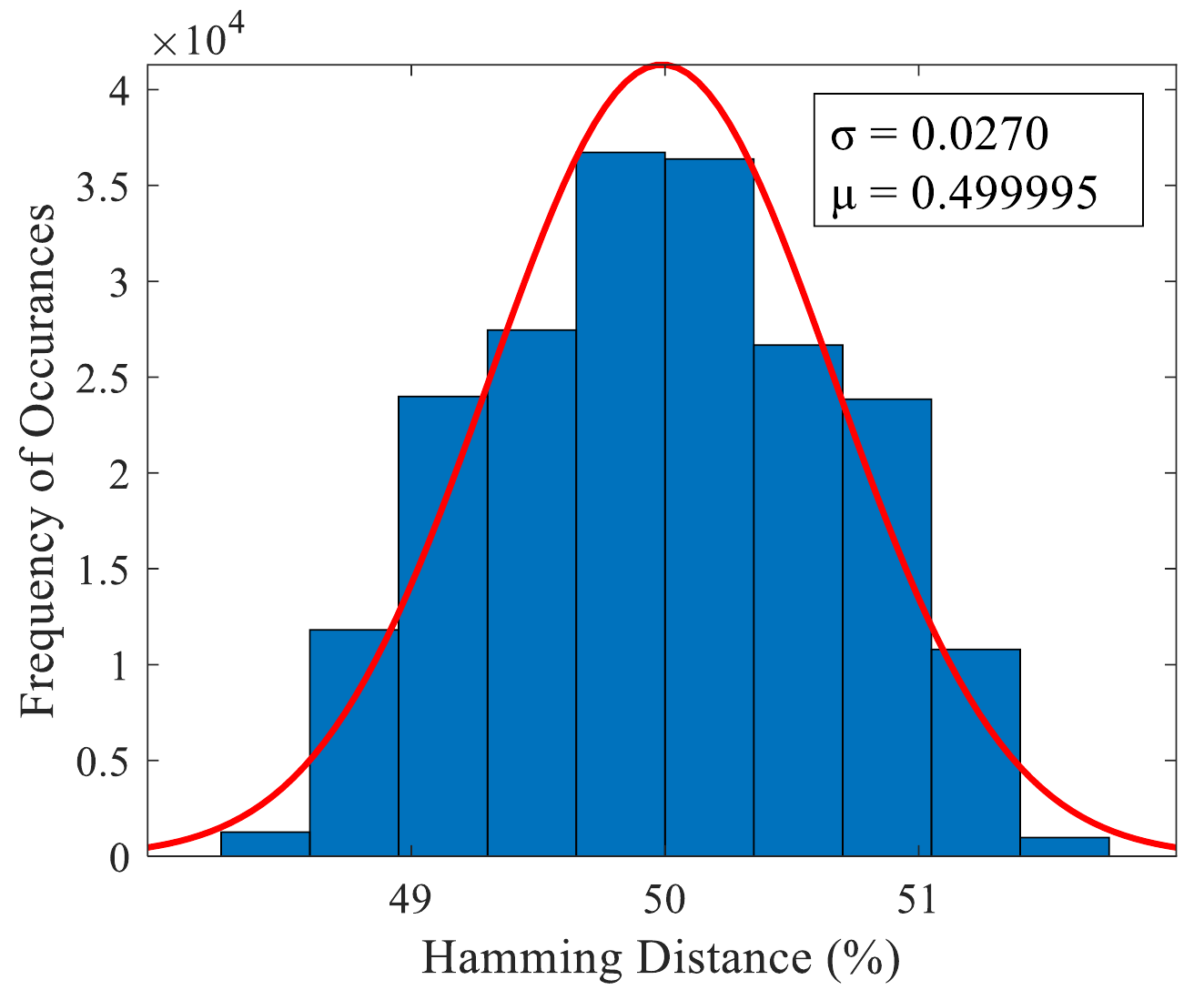}
		\caption{Multikey-PUF 1}
		\label{fig:uniqueness2}
	\end{subfigure}
	~ 
	\begin{subfigure}[b]{0.3\textwidth}
		\includegraphics[width=\textwidth]{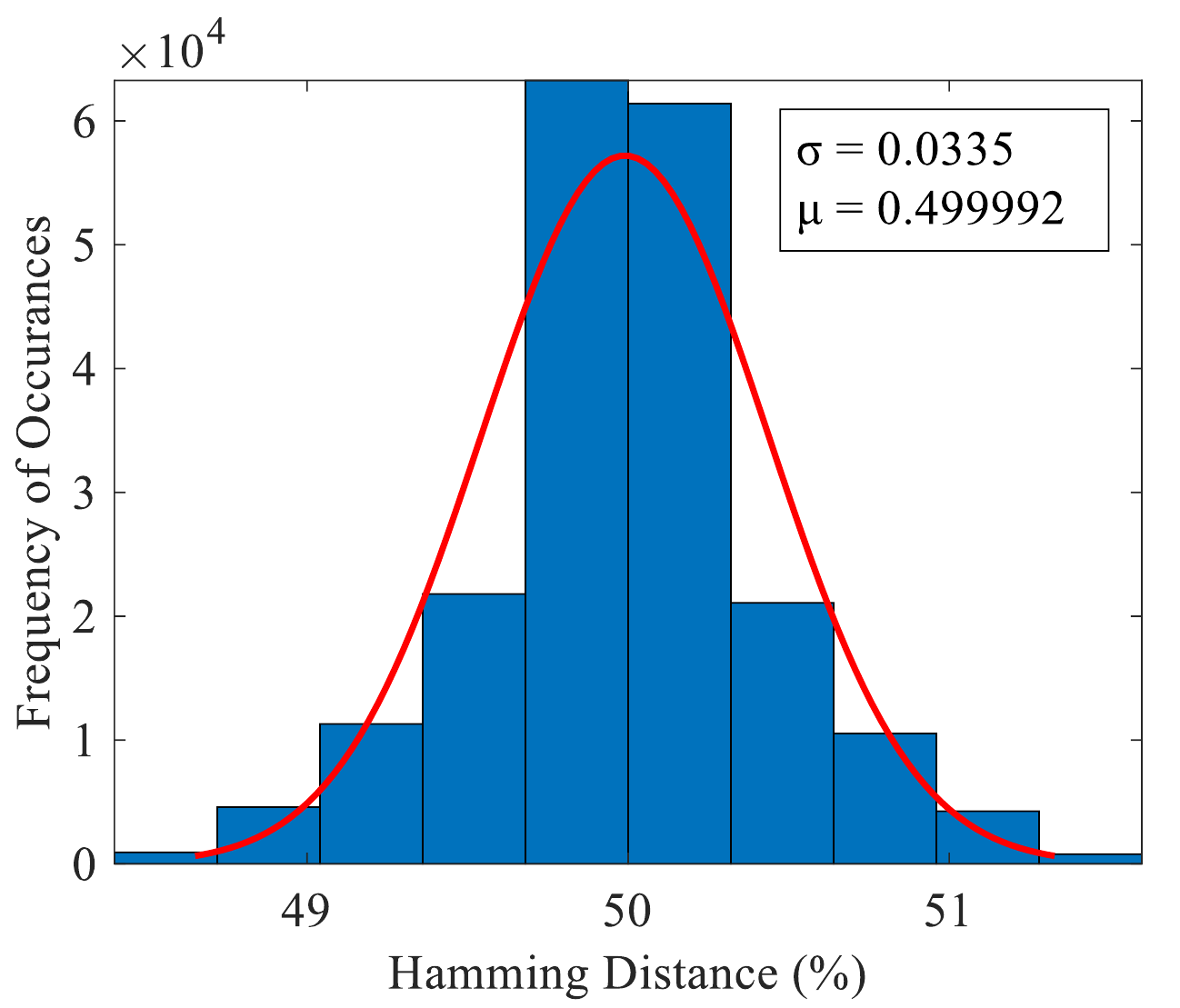}
		\caption{Multikey-PUF 2}
		\label{fig:uniqueness3}
	\end{subfigure}
	~ 
	\begin{subfigure}[b]{0.3\textwidth}
		\includegraphics[width=\textwidth]{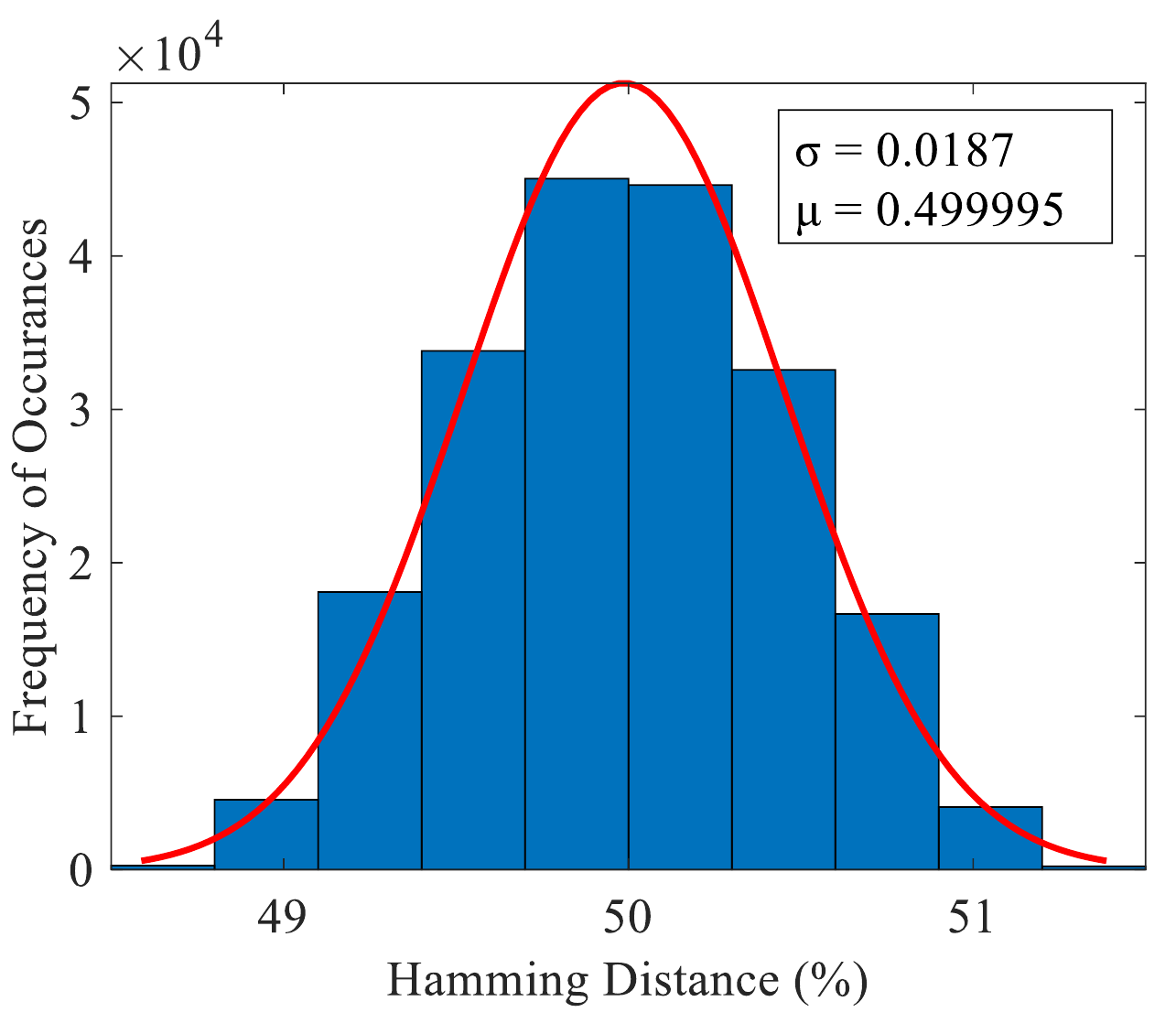}
		\caption{Multikey-PUF 3}
		\label{fig:uniqueness4}
	\end{subfigure}
	
	\caption{Uniqueness of Multikey-PUF deployed on multiple FPGAs}
	\label{FIG:Uniqueness}
\end{figure*}

\begin{figure*}
	\centering
	\begin{subfigure}[b]{0.3\textwidth}
		\includegraphics[width=\textwidth]{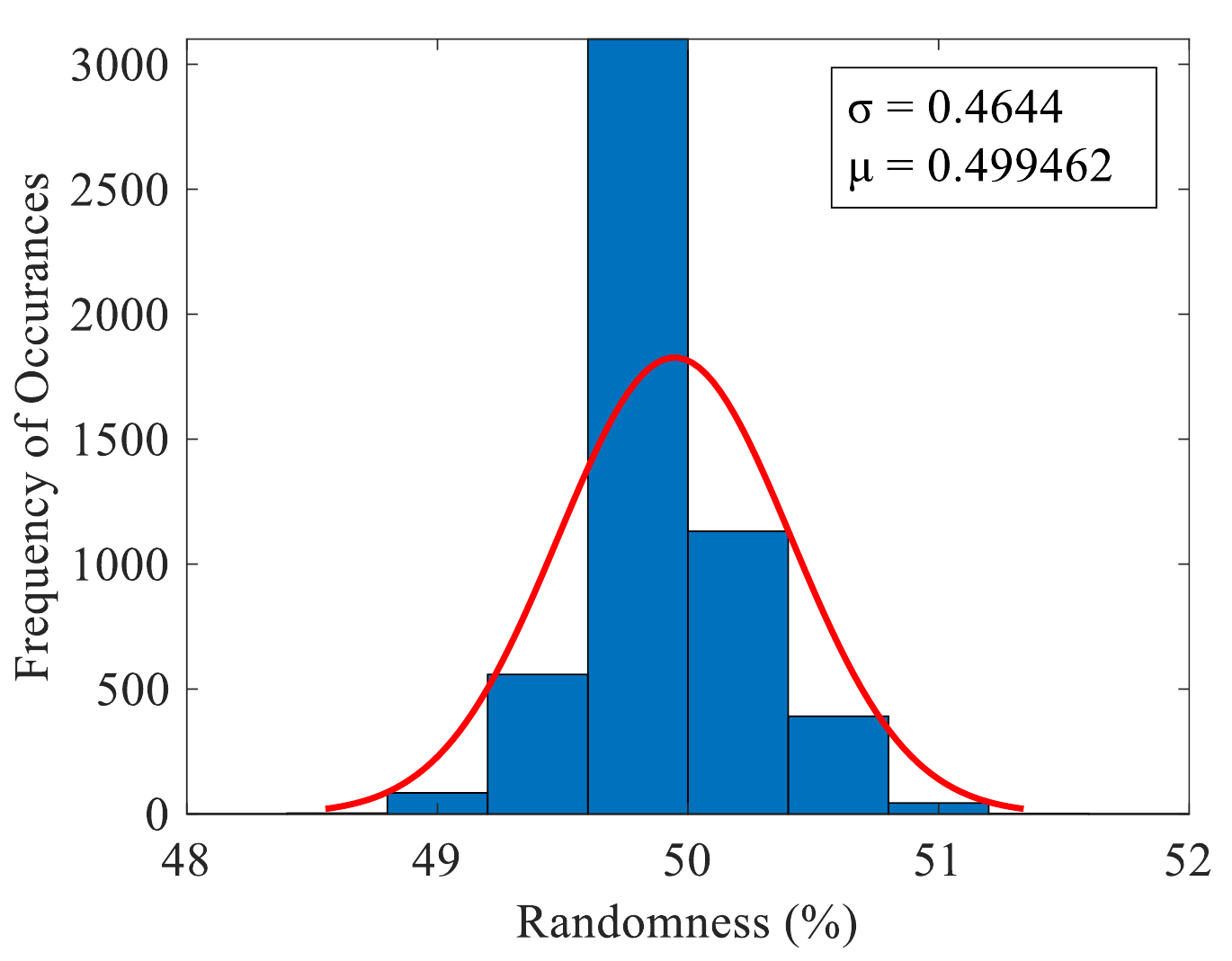}
		\caption{Multikey-PUF 1}
		\label{fig:randomness1}
	\end{subfigure}
	~ 
	\begin{subfigure}[b]{0.3\textwidth}
		\includegraphics[width=\textwidth]{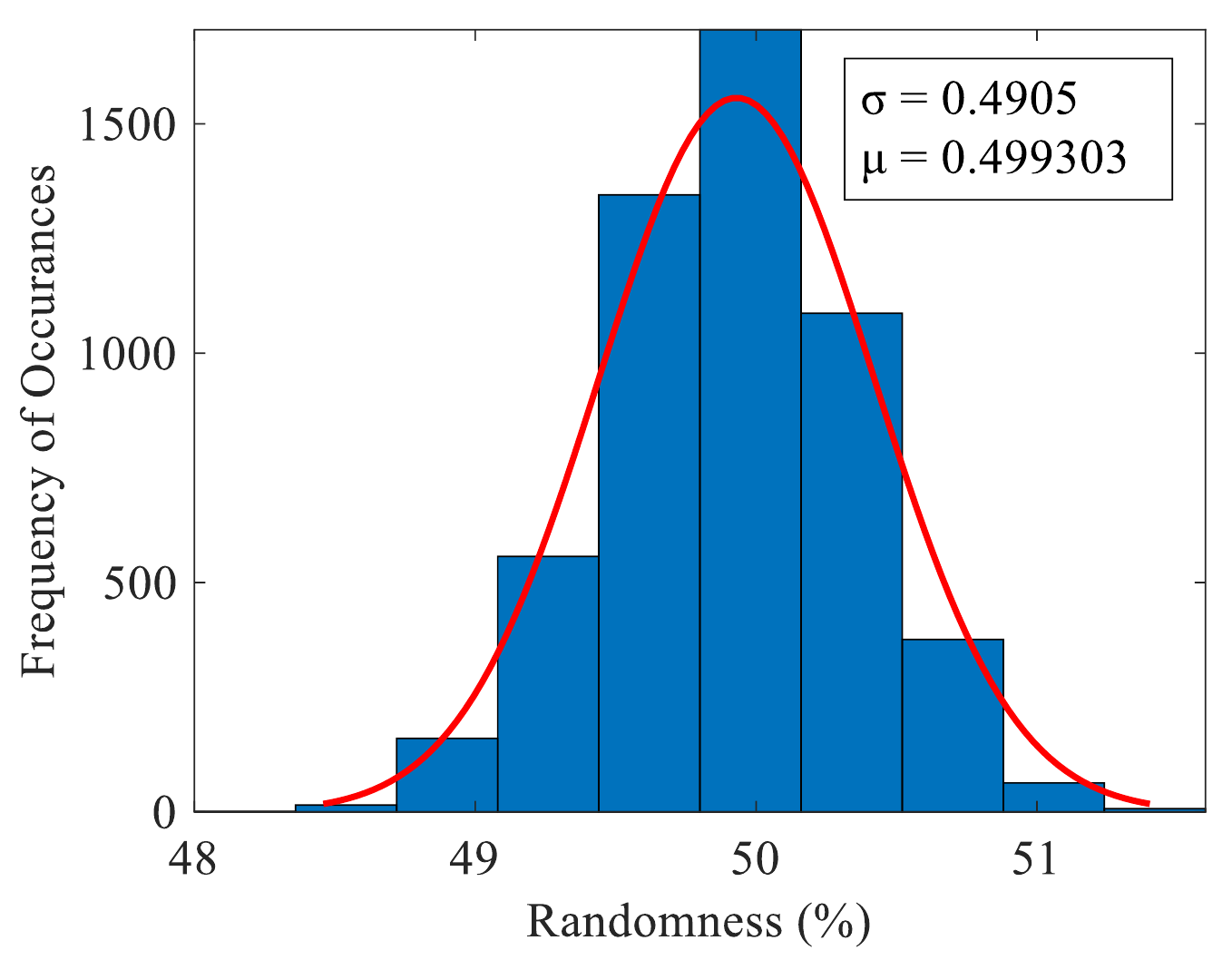}
		\caption{Multikey-PUF 2}
		\label{fig:randomness2}
	\end{subfigure}
	~ 
	\begin{subfigure}[b]{0.3\textwidth}
		\includegraphics[width=\textwidth]{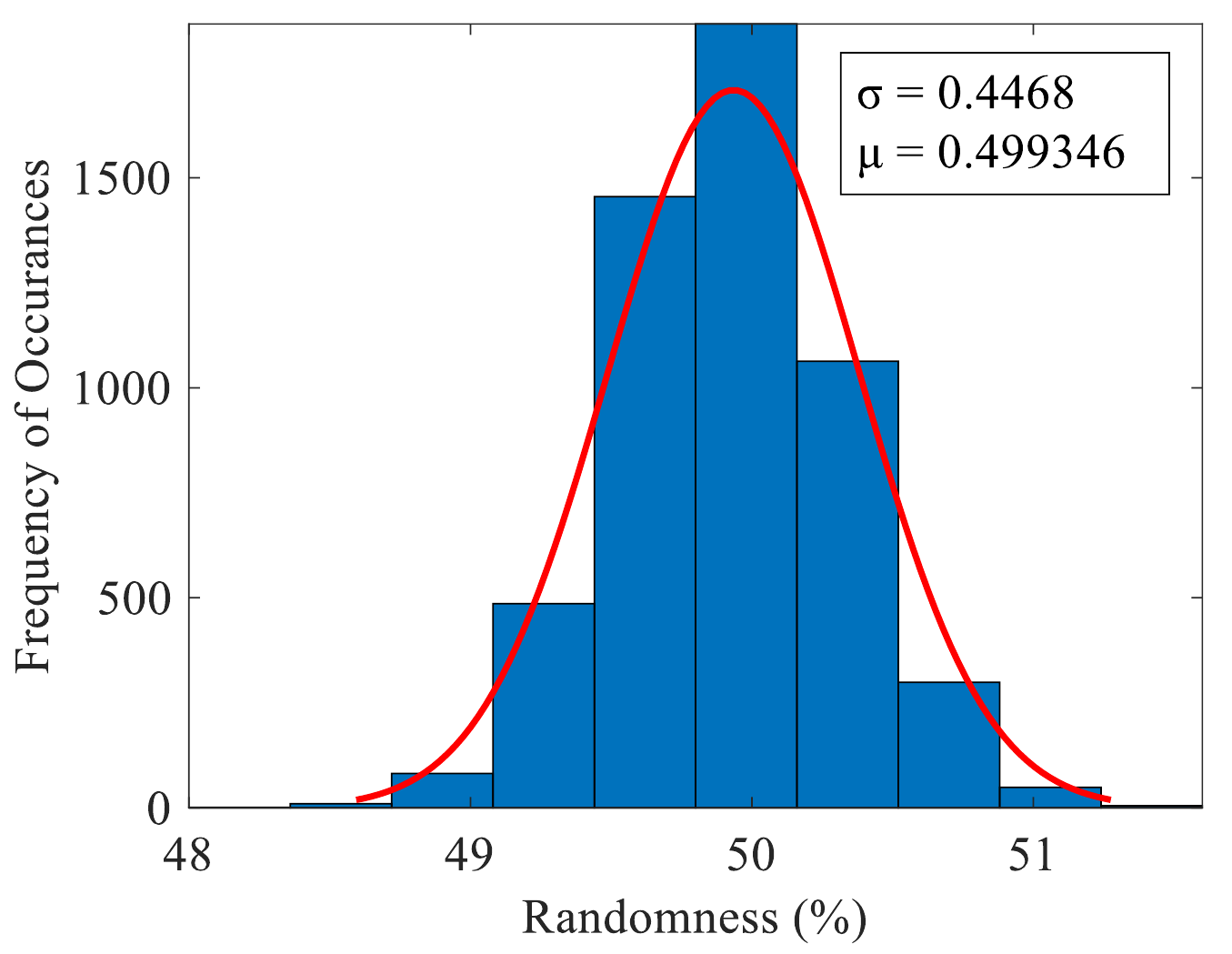}
		\caption{Multikey-PUF 3}
		\label{fig:randomness3}
	\end{subfigure}
	
	\caption{Randomness of Multikey-PUF deployed on multiple FPGAs}
	\label{FIG:Randomness}
\end{figure*}
The other Board that was used for deploying PUF was the Basys-3 FPGA board with a Xilinx FPGA on the board. Three different PUF instances were developed on the Basys Board. The PUF is evaluated through three Figures of Merit (FoMs), Uniqueness, Randomness, and Reliability. A Silicon PUF is vulnerable to the ambient temperature in many applications. Hence, a temperature variation analysis of the FoMs was also performed

\subsection {Uniqueness and Reliability}		


The ability of PUF to generate a unique key is called "Uniqueness". PUF uses the manufacturing variations introduced during the fabrication of the circuit to generate the outputs. Uniqueness can be calculated using hamming distance between the keys generated by the PUF. The ideal value of uniqueness for the keys to be used for cryptographic purposes is 50 \%. Fig. \ref{FIG:Uniqueness} shows the uniqueness of multiple PUF modules instantiated on two FPGAs.

Two PUF modules were instantiated on the Arduino MKR Vidor 4000 FPGA and a Basys 3 FPGA development board was instantiated with a PUF module. As shown in Fig. \ref{FIG:Uniqueness}, the PUF modules have a mean hamming distance of around 50 \% with a minimal standard deviation. To calculate the uniqueness, the keys for each module were collected over a period of 1 hour and a total of 200,000 keys were collected from each module. There were no collisions between the keys during the time the keys were collected. 

The reliability of PUF is the ability of the module to generate keys at a reliable rate. In general, a PUF module gives responses as outputs. A PUF module is expected to generate the same response for a respective challenge under various circumstances, such as temperature variations, power supply variations, and aging effects. The design of the Multikey PUF generates a different response for each run. Hence the reliability of the presented design lies in generating a new key every time the module is run. In this case, the reliability can be determined using hamming distance. 200,000 keys were generated over the period of 1 hour. The reliability can also be determined from Fig. \ref{FIG:Uniqueness}. The hamming distance is around the ideal value, 50 \% which shows a reliable functioning of the module.


\begin{figure*}
	\centering
	\begin{subfigure}[b]{0.3\textwidth}
		\includegraphics[width=\textwidth]{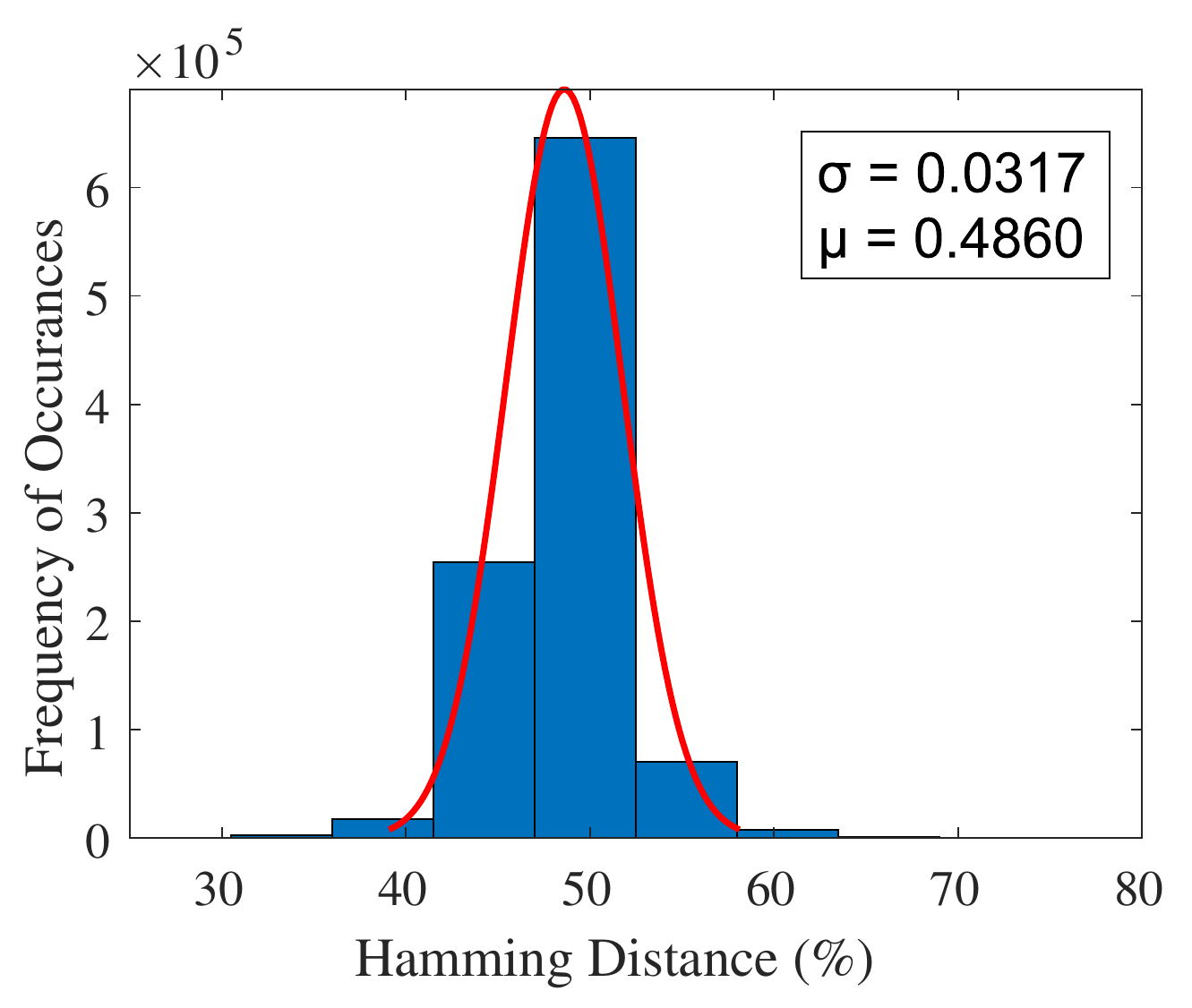}
		\caption{20\textdegree F}
		\label{fig:h20F}
	\end{subfigure}
	~ 
	\begin{subfigure}[b]{0.3\textwidth}
		\includegraphics[width=\textwidth]{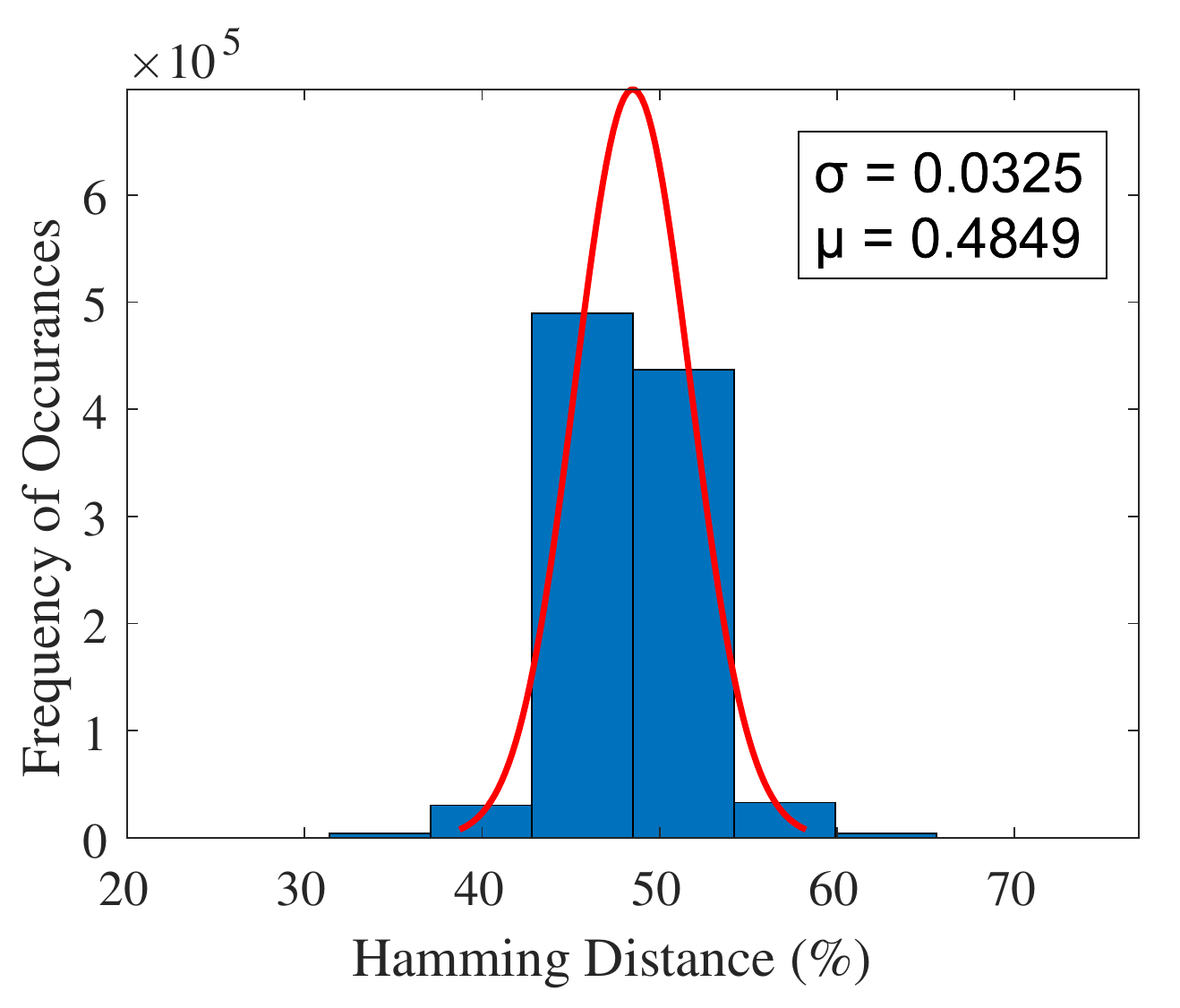}
		\caption{40\textdegree F}
		\label{fig:h40F}
	\end{subfigure}
	~ 
	\begin{subfigure}[b]{0.3\textwidth}
		\includegraphics[width=\textwidth]{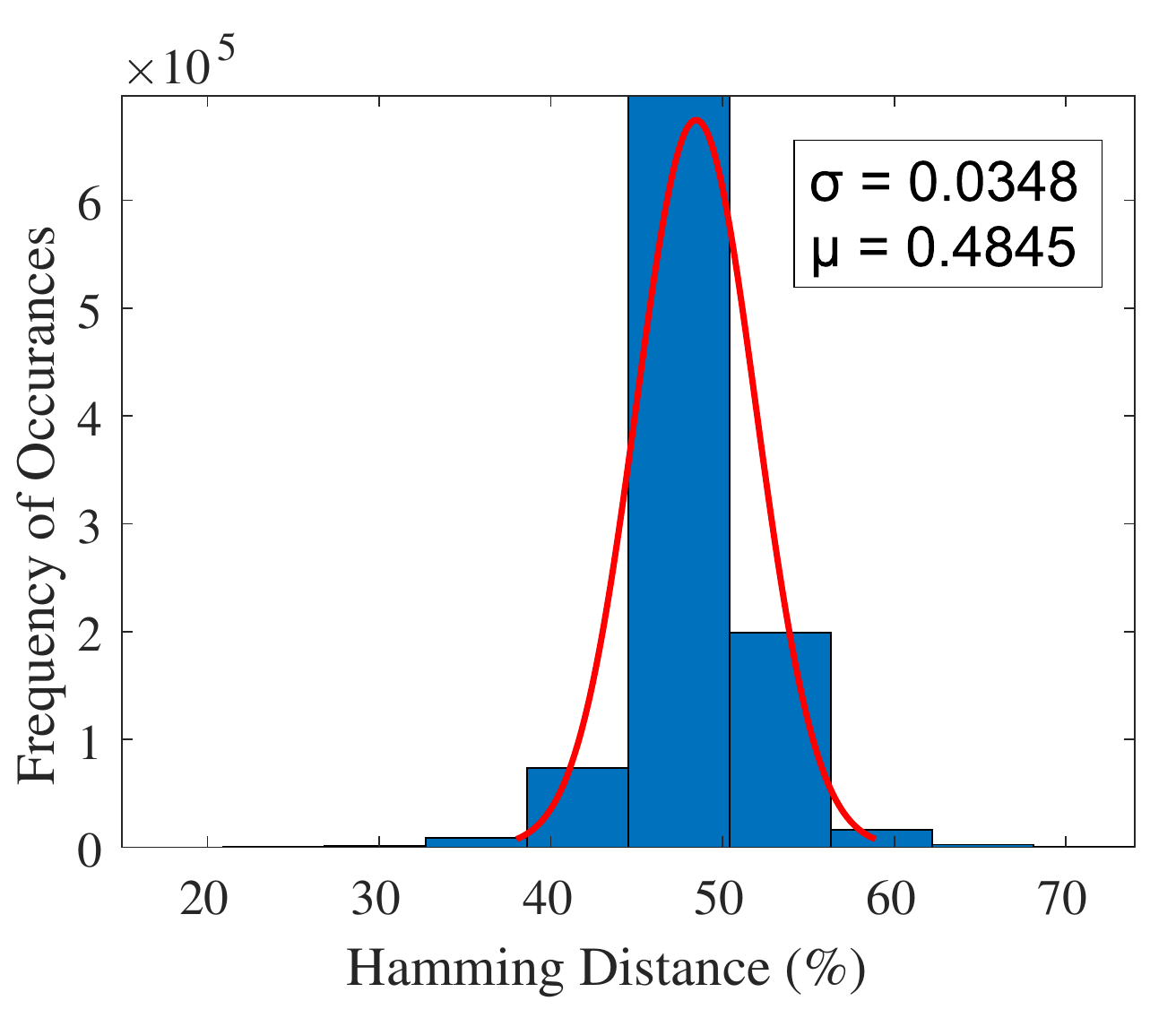}
		\caption{60\textdegree F}
		\label{fig:h60F}
	\end{subfigure}
	~ 
	\begin{subfigure}[b]{0.3\textwidth}
		\includegraphics[width=\textwidth]{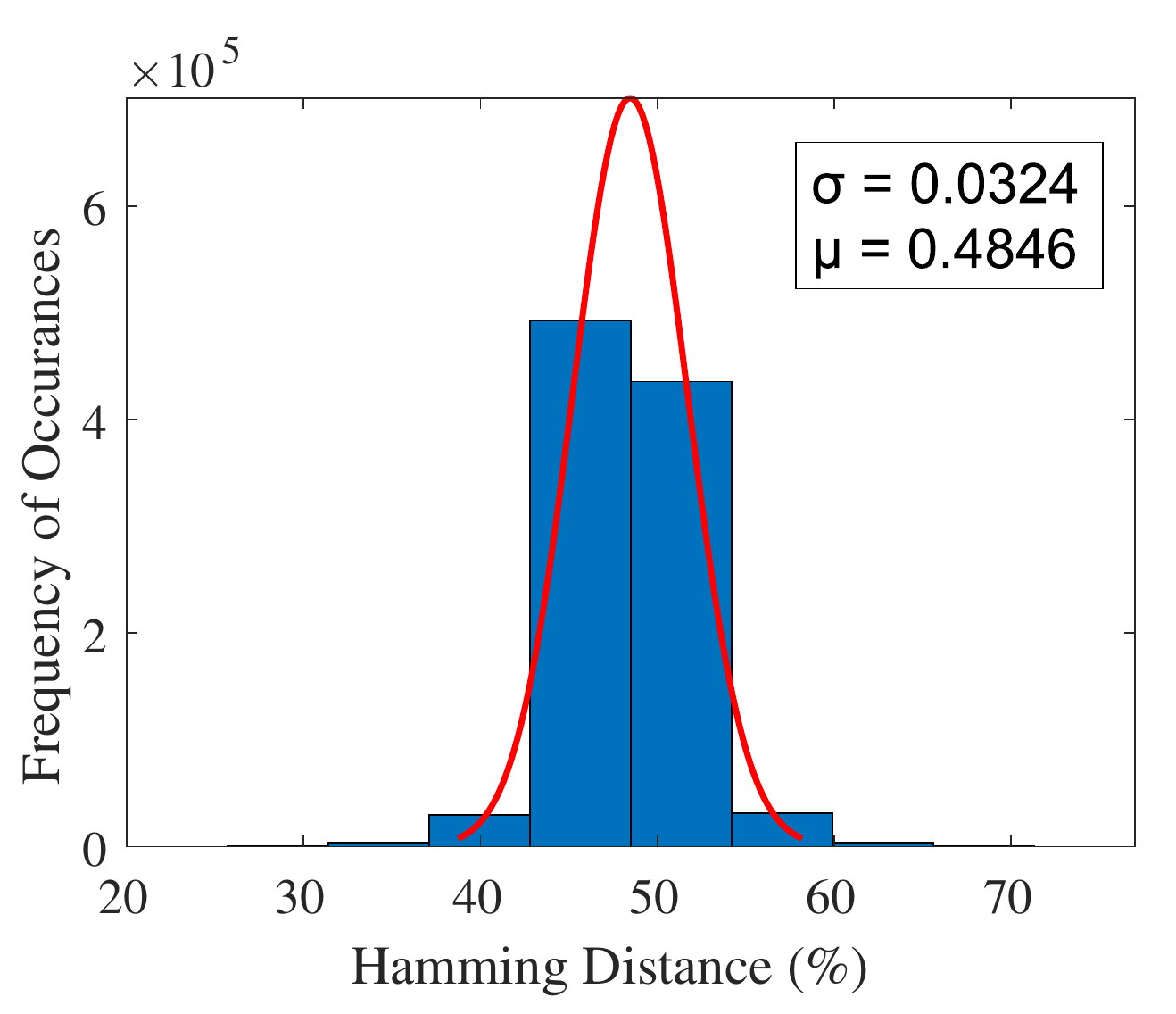}
		\caption{80\textdegree F}
		\label{fig:h80F}
	\end{subfigure}
	~ 
	\begin{subfigure}[b]{0.3\textwidth}
		\includegraphics[width=\textwidth]{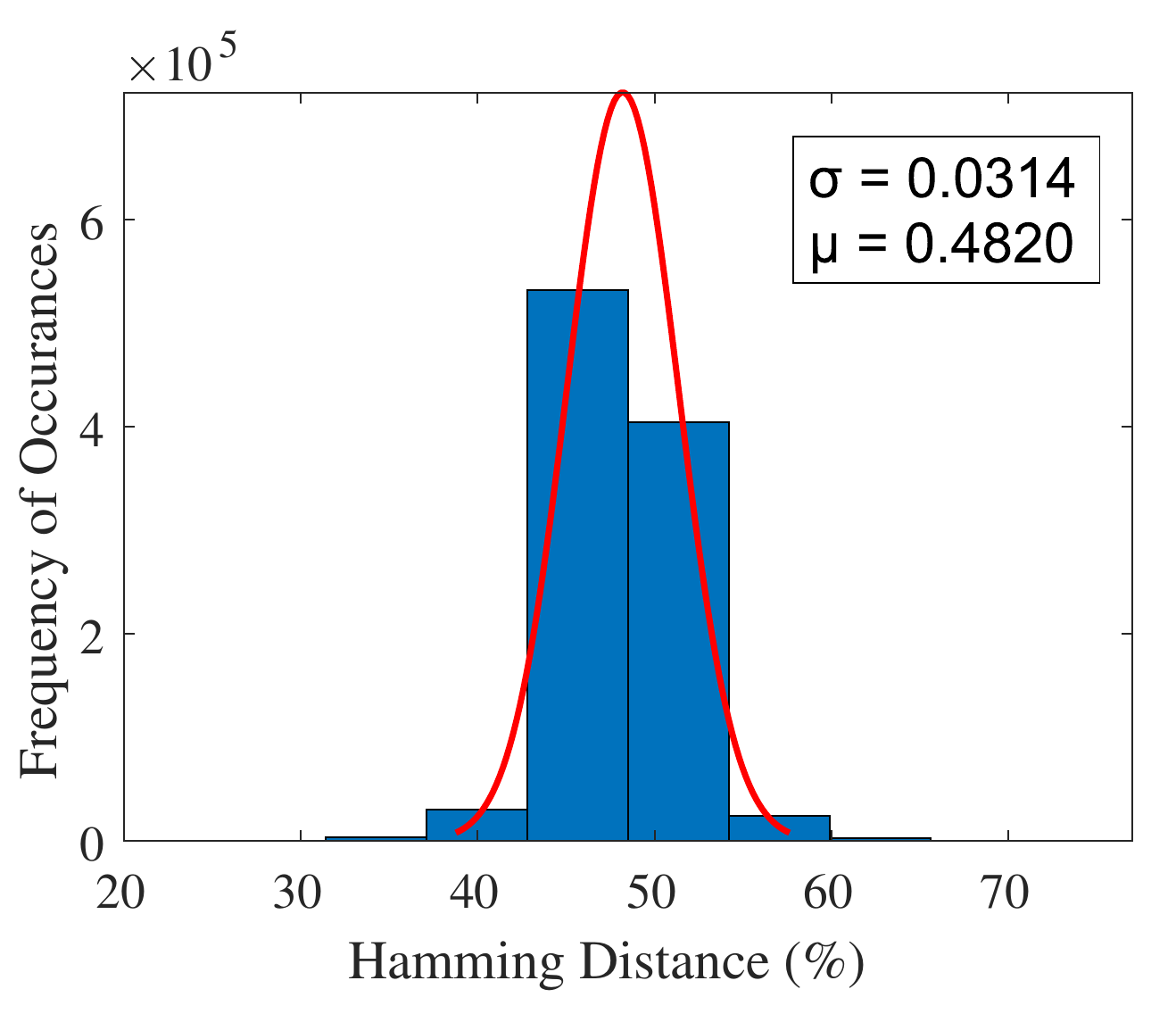}
		\caption{120\textdegree F}
		\label{fig:h120F}
	\end{subfigure}
	~ 
	\begin{subfigure}[b]{0.3\textwidth}
		\includegraphics[width=\textwidth]{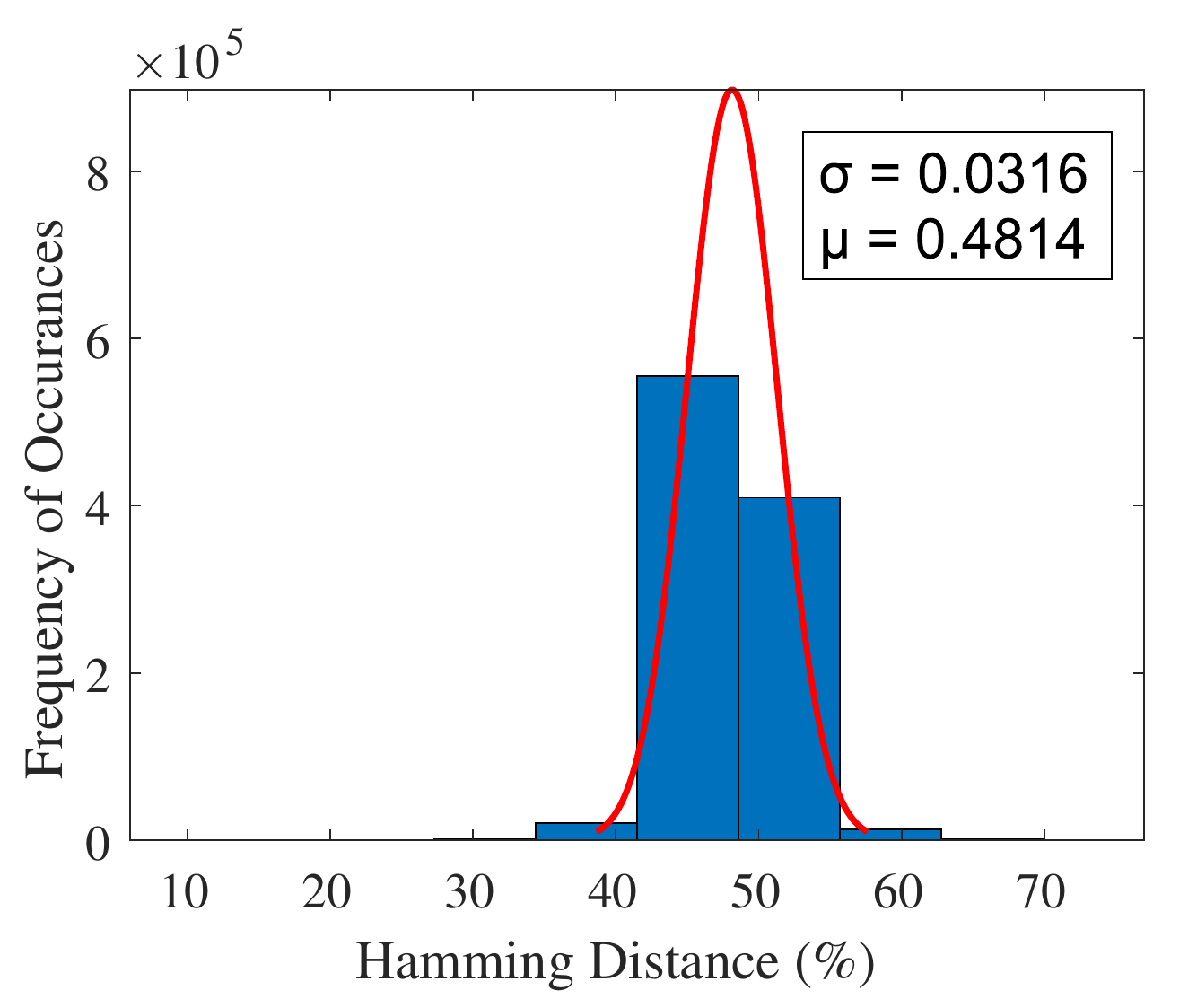}
		\caption{140\textdegree F}
		\label{fig:h140F}
	\end{subfigure}
	
	\caption{Uniqueness of PUF Modules at Different Temperatures (a. 20\textdegree F b. 40\textdegree F c. 60\textdegree F d. 80\textdegree F e. 120\textdegree F f. 140\textdegree F)}
	\label{FIG:Uniqueness-Temp}
\end{figure*}	
\begin{figure*}
	\centering
	\begin{subfigure}[b]{0.3\textwidth}
		\includegraphics[width=\textwidth]{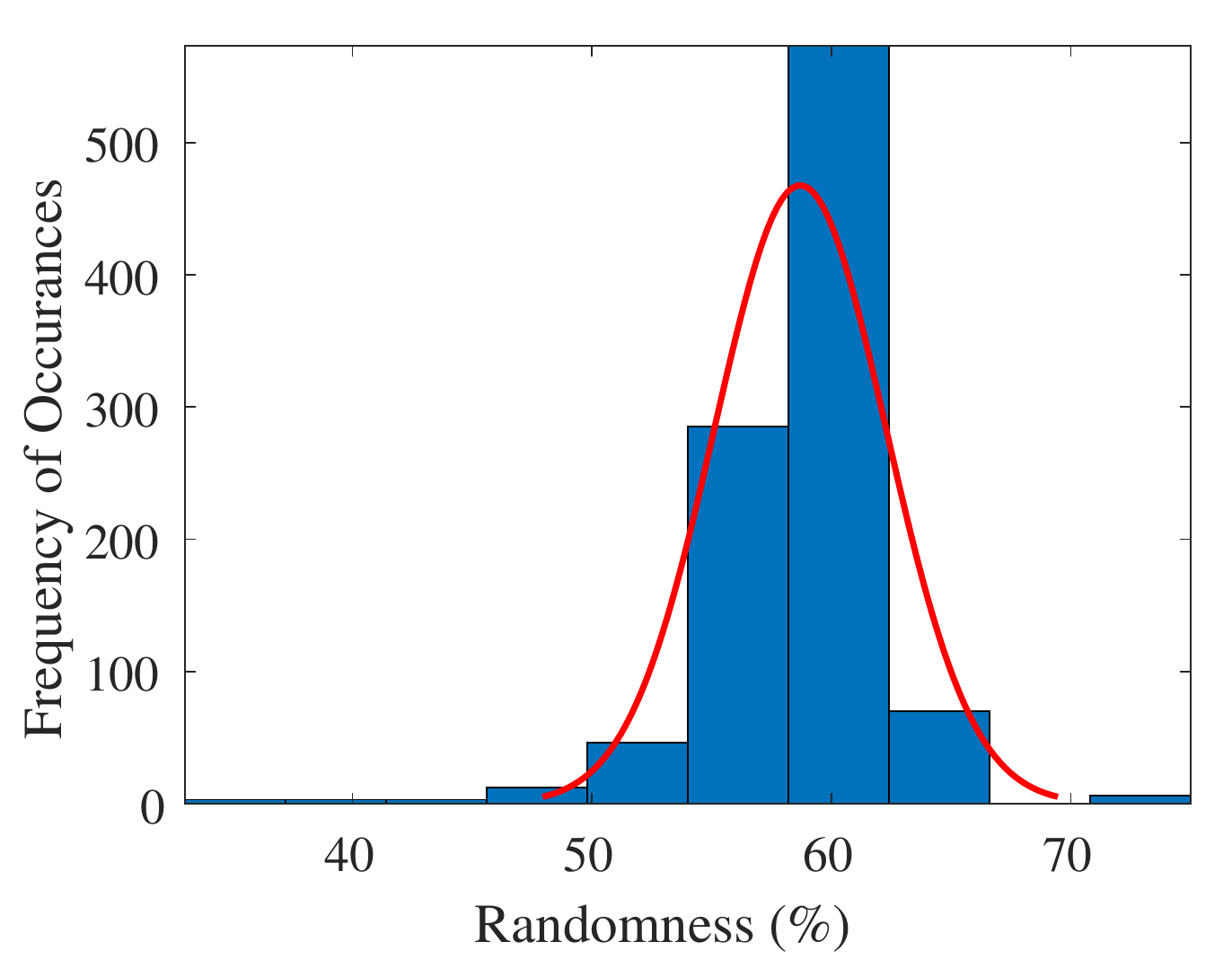}
		\caption{20\textdegree F}
		\label{fig:20F}
	\end{subfigure}
	~ 
	\begin{subfigure}[b]{0.3\textwidth}
		\includegraphics[width=\textwidth]{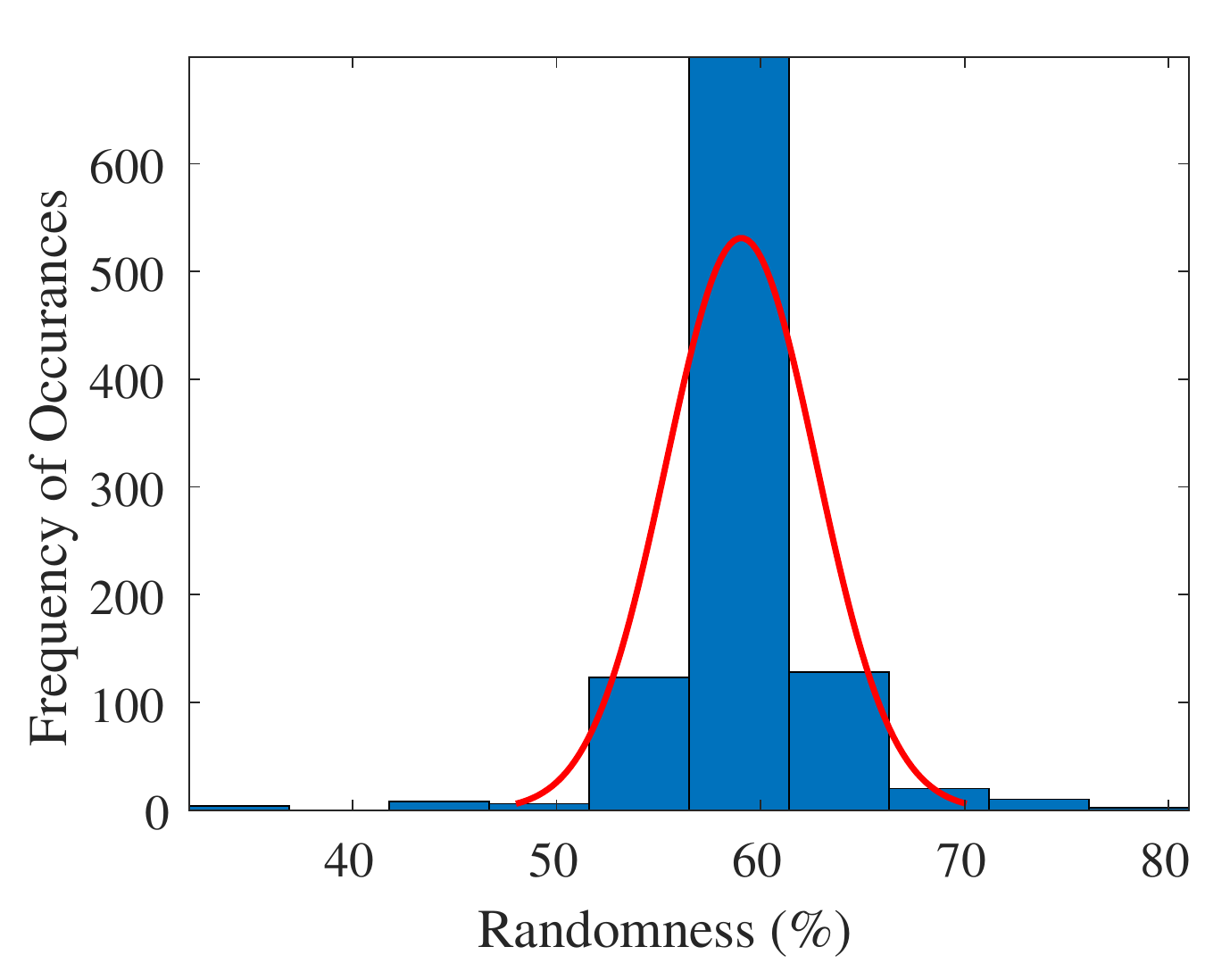}
		\caption{40\textdegree F}
		\label{fig:}
	\end{subfigure}
	~ 
	\begin{subfigure}[b]{0.3\textwidth}
		\includegraphics[width=\textwidth]{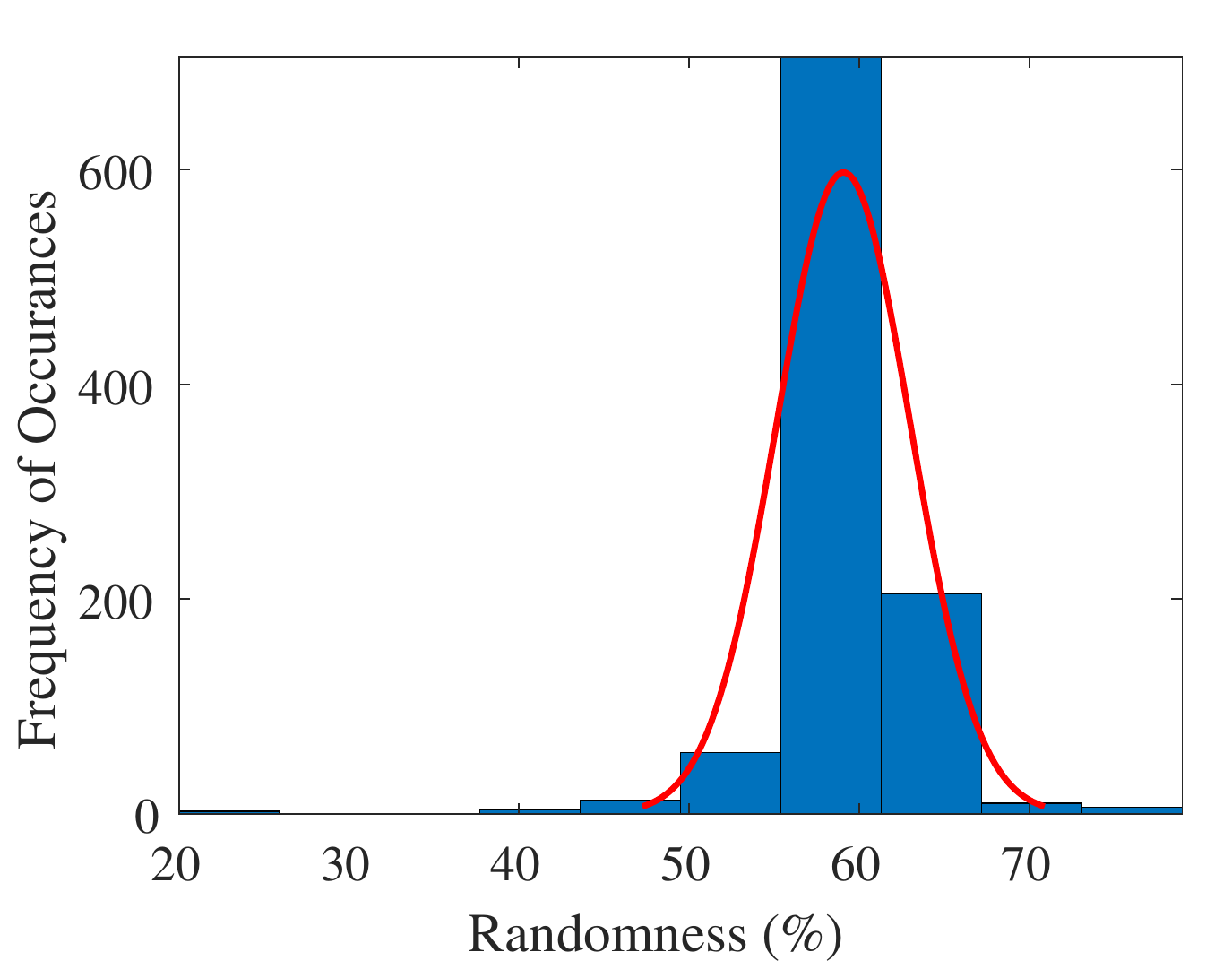}
		\caption{60\textdegree F}
		\label{fig:}
	\end{subfigure}
	~ 
	\begin{subfigure}[b]{0.3\textwidth}
		\includegraphics[width=\textwidth]{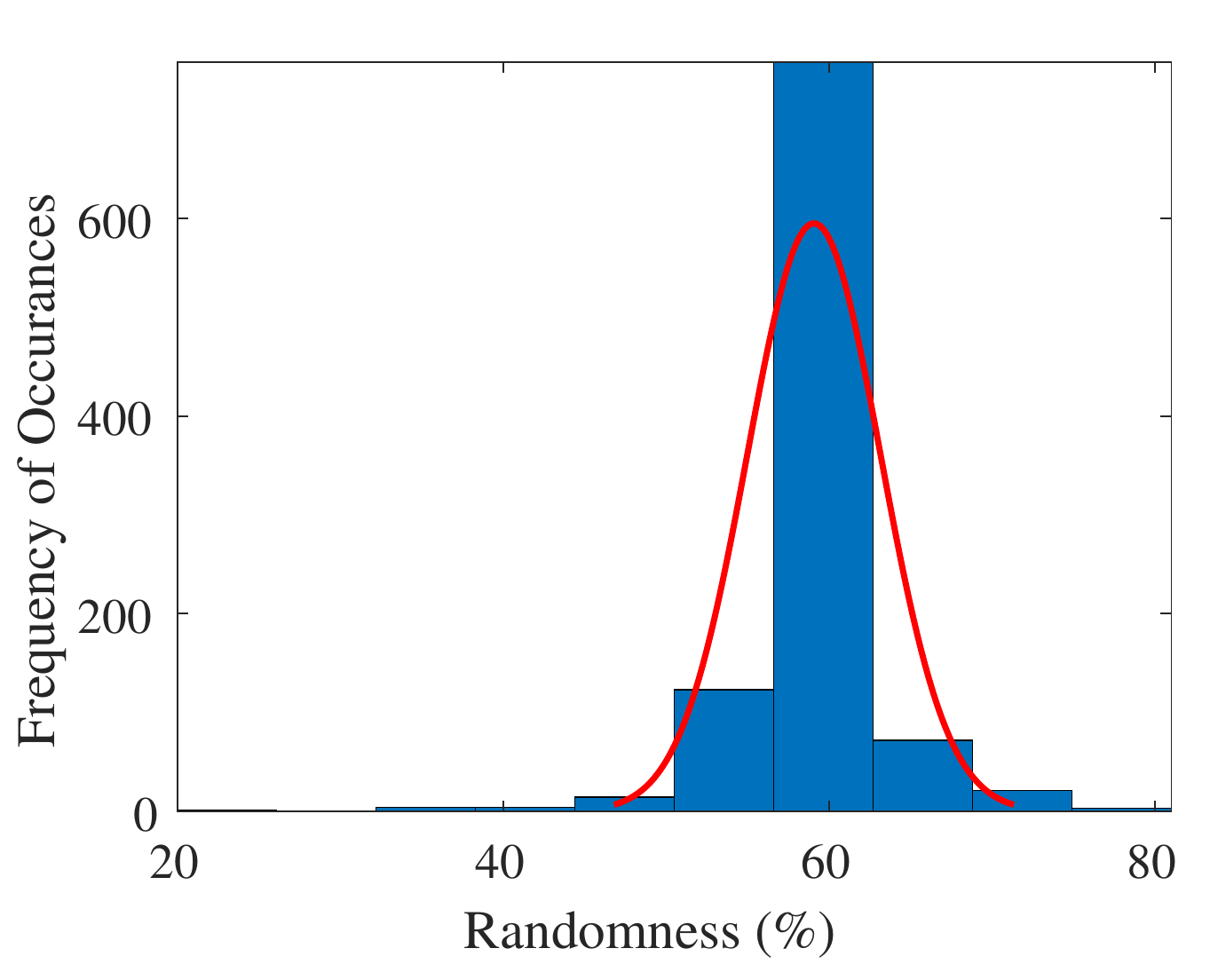}
		\caption{80\textdegree F}
		\label{fig:}
	\end{subfigure}
	~ 
	\begin{subfigure}[b]{0.3\textwidth}
		\includegraphics[width=\textwidth]{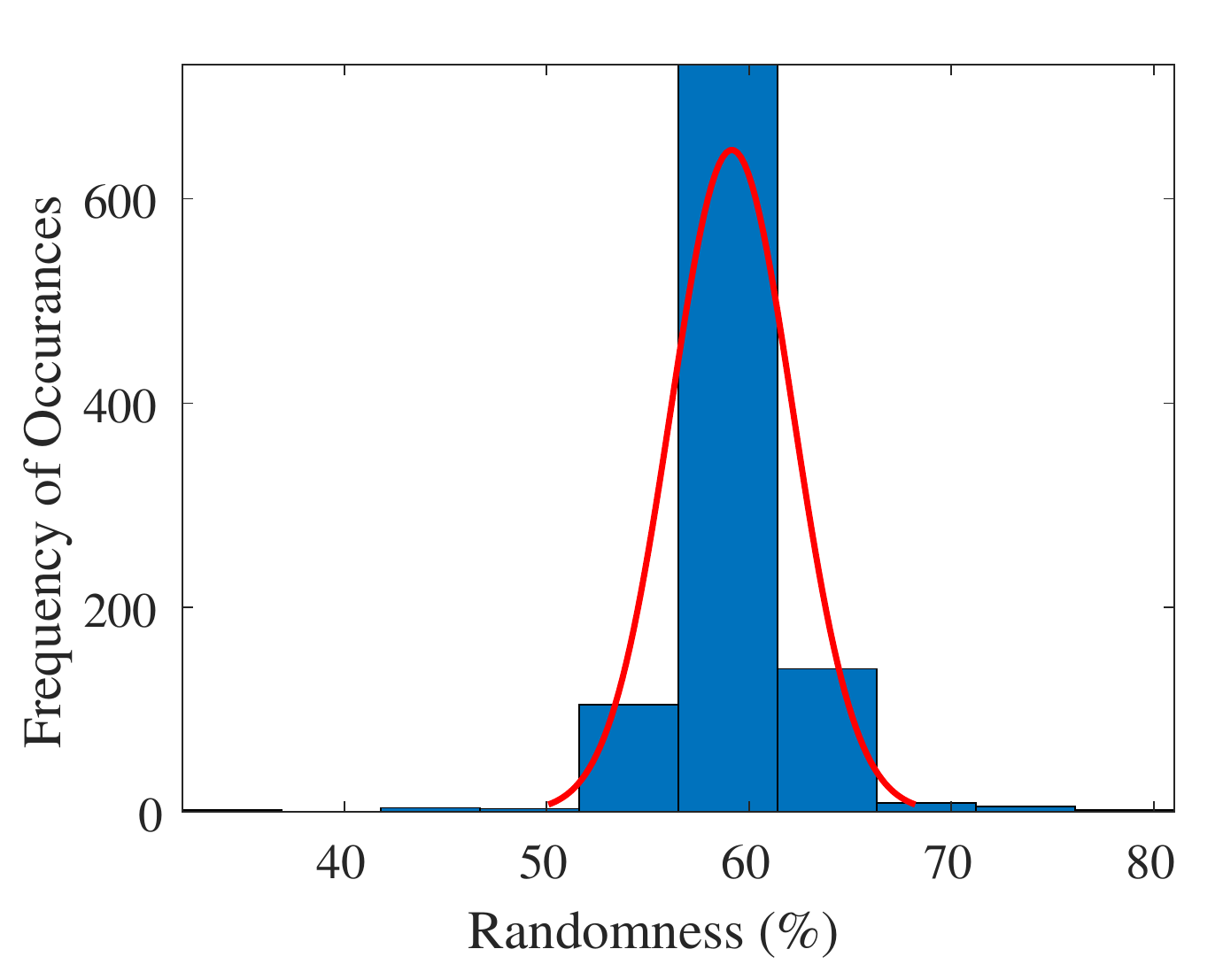}
		\caption{120\textdegree F}
		\label{fig:}
	\end{subfigure}
	~ 
	\begin{subfigure}[b]{0.3\textwidth}
		\includegraphics[width=\textwidth]{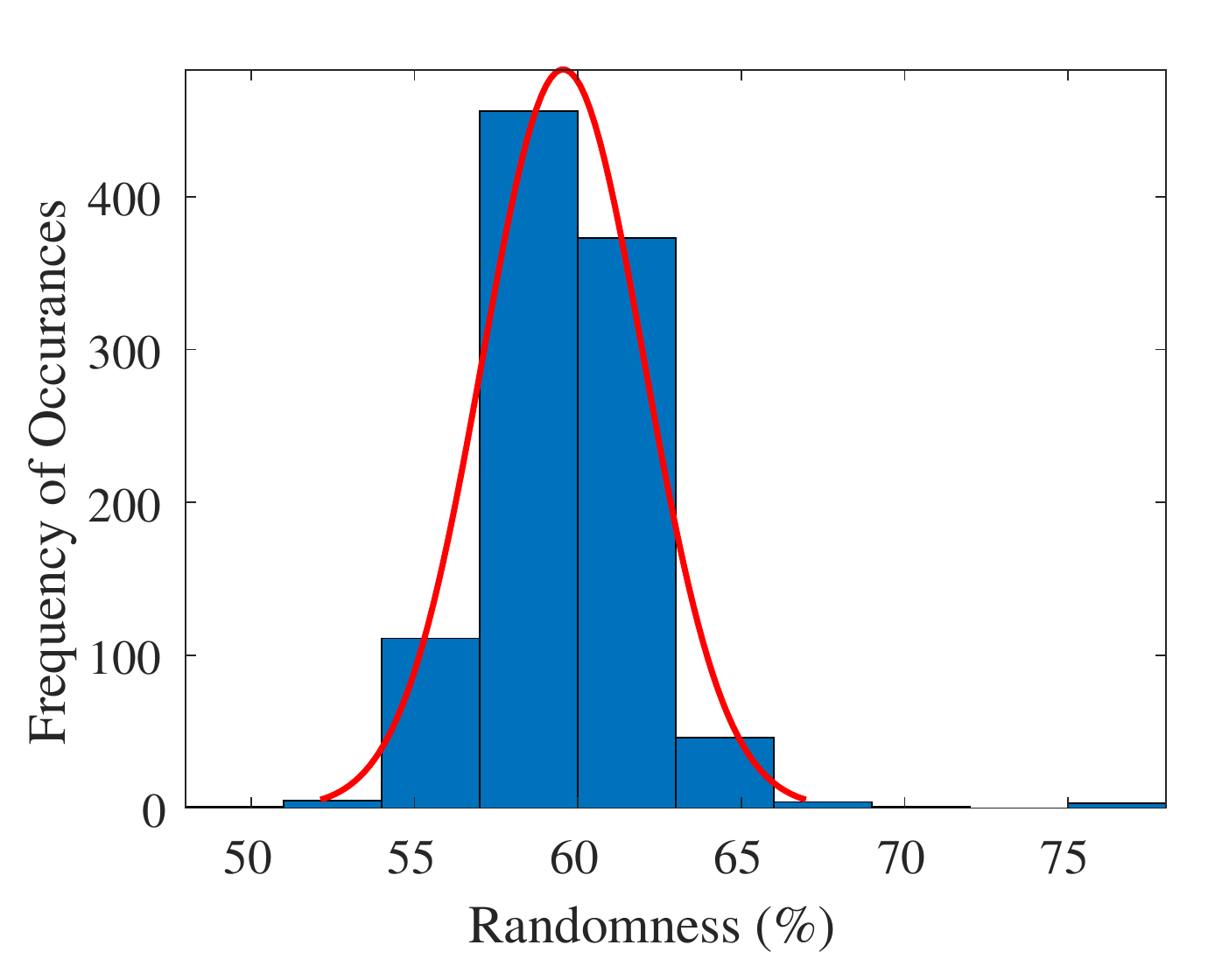}
		\caption{140\textdegree F}
		\label{fig:}
	\end{subfigure}
	
	\caption{Randomness of PUF Modules at Different Temperatures (a. 20\textdegree F b. 40\textdegree F c. 60\textdegree F d. 80\textdegree F e. 120\textdegree F f. 140\textdegree F)}
	\label{FIG:Randomness-Temp}
\end{figure*}

\subsection{Randomness}

Randomness determines the distribution of bits 1 and 0 in a key generated by the PUF module. A key, to be used for cryptographic purposes has to have an equal distribution of 1 and 0 in the output binary bits. The ideal value of randomness in PUF output is 50 \%. 256-bit keys were generated from the PUF modules to test the randomness of the keys. Fig. \ref{FIG:Randomness} shows the randomness of the output keys generated by the PUF modules instantiated on two FPGAs. As shown in the figure, the randomness of the PUF modules is about 50 \% which is close to the ideal value expected. 


The essential properties (repeatability, randomness, long period, Insensitive to seeds) of RNG which helps to make the IoT devices secured. Moreover making a good random number has some more properties to consider which are portability, efficiency, disjoint sub-sequences, and homogeneity.

\subsection{Temperature Analysis of Multikey-PUF}


Silicon PUFs are vulnerable to temperature variations significantly. An error introduced into a PUF can significantly affect the output bit generation by introducing errors into the system. This can prove to be a devastating outcome in many applications. Hence the PUF modules were tested under the operating temperatures of 20\textsuperscript{o}F, 40\textsuperscript{o}F, 60\textsuperscript{o}F, 80\textsuperscript{o}F, 120\textsuperscript{o}F, 140\textsuperscript{o}F, respectively. The uniqueness, randomness, and reliability of PUF were evaluated during the temperature variations.

Fig. \ref{FIG:Uniqueness-Temp} shows the uniqueness and reliability of PUF under temperature variations. The output keys of PUF have always shown a hamming distance of 50 \%, which is close to the ideal value. Fig. \ref{FIG:Randomness-Temp} shows the randomness of PUF under temperature variations. The output of PUF is also near the ideal value.

Table 1 depicts the prototype's characteristics. The key is generated in 1 second on average. The length of time is determined by the PUF design. The same key is utilized for the current session after the keys have been produced. On the Arduino board, there is an FPGA. The board consumes 58 mW of electricity on average. The average PUF Uniqueness for six different temperatures is 48.39\% and the average reliability is almost 0\%. Moreover, the average PUF randomness is 59.0994\%.

\begin{table}[h!]
	\centering
	\begin{tabular}{|c | c|} 
		\hline
		Server & Device and Values\\ [0.5ex] 
		\hline
		IoMT Device & Arduino MKR\\ & VIDOR 4000 \\ 
		\hline
		PUF & FPGA \\
		\hline
		Avg. time to generate the key & 1 s\\
		\hline
		Avg. power consumed & 0.58 W\\
		\hline
		PUF Uniqueness for various temperature  & \\
		20\textdegree F & 48.60\% \\ 
		40\textdegree F & 48.49\%\\ 
		60\textdegree F & 48.45\%\\ 
		80\textdegree F & 48.46\%\\ 
		120\textdegree F & 48.20\%\\ 
		140\textdegree F & 48.14\%\\ 
		\hline
		Avg. PUF Reliability for various temperature  & Almost 0\%\\
		\hline
		PUF Randomness for various temperature  & \%\\
		20\textdegree F & 58.7003\% \\ 
		40\textdegree F & 59.0492\%\\ 
		60\textdegree F & 59.0800\%\\ 
		80\textdegree F & 59.0688\%\\ 
		120\textdegree F & 59.1483\%\\ 
		140\textdegree F & 59.5500\%\\ 
		\hline
	\end{tabular}
	\caption{Characteristics of the System}
	\label{table:1}
\end{table}

\section{Conclusion and Future work}
The most important concern with intelligent connected devices is security. PUFs are an interesting new technology that can be used to generate keys for our devices' security and identification. This paper focuses on Hybrid-Multikey-PUF and their applicability in generating secure IoT systems through encryption key generation. In general Silicon, PUFs are vulnerable to the ambient temperature though we have considered different temperatures (20\textdegree F, 40\textdegree F, 60\textdegree F, 80\textdegree F, 120\textdegree F, 140\textdegree ) to evaluate three Figures of Merit (FoMs), Uniqueness, Randomness, and Reliability. We obtained nearly optimal values for the three FoMs of PUFs, and multikey-PUF might be used in IoT devices for security purposes in various temperatures. The future work of the presented results includes analyzing the output keys using various test suits and developing an application targeted toward the Internet of Things and resource-constrained devices using the Hybrid-Multikey-PUF. Also, we would like to extend this work to ultra low power IoT environments.


\bibliographystyle{IEEEtran}
\balance 
\bibliography{Bibliography}

\begin{thebibliography}{10}
\providecommand{\url}[1]{#1}
\csname url@samestyle\endcsname
\providecommand{\newblock}{\relax}
\providecommand{\bibinfo}[2]{#2}
\providecommand{\BIBentrySTDinterwordspacing}{\spaceskip=0pt\relax}
\providecommand{\BIBentryALTinterwordstretchfactor}{4}
\providecommand{\BIBentryALTinterwordspacing}{\spaceskip=\fontdimen2\font plus
\BIBentryALTinterwordstretchfactor\fontdimen3\font minus
  \fontdimen4\font\relax}
\providecommand{\BIBforeignlanguage}[2]{{%
\expandafter\ifx\csname l@#1\endcsname\relax
\typeout{** WARNING: IEEEtran.bst: No hyphenation pattern has been}%
\typeout{** loaded for the language `#1'. Using the pattern for}%
\typeout{** the default language instead.}%
\else
\language=\csname l@#1\endcsname
\fi
#2}}
\providecommand{\BIBdecl}{\relax}
\BIBdecl

\bibitem{bakiri2018survey}
M.~Bakiri, C.~Guyeux, J.-F. Couchot, and A.~K. Oudjida, ``Survey on hardware
  implementation of random number generators on fpga: Theory and experimental
  analyses,'' \emph{Computer Science Review}, vol.~27, pp. 135--153, 2018.

\bibitem{Sathya2021}
\BIBentryALTinterwordspacing
K.~Sathya, J.~Premalatha, and V.~Rajasekar, ``Investigation of strength and
  security of pseudo random number generators,'' \emph{{IOP} Conference Series:
  Materials Science and Engineering}, vol. 1055, no.~1, p. 012076, feb 2021.
  [Online]. Available: \url{https://doi.org/10.1088/1757-899x/1055/1/012076}
\BIBentrySTDinterwordspacing

\bibitem{adewopo2022review}
V.~Adewopo, N.~Elsayed, Z.~ElSayed, M.~Ozer, A.~Abdelgawad, and M.~Bayoumi,
  ``Review on action recognition for accident detection in smart city
  transportation systems,'' \emph{arXiv e-prints}, pp. arXiv--2208, 2022.

\bibitem{9595288}
M.~I. Mahmud, A.~Abdelgawad, V.~P. Yanambaka, and K.~Yelamarthi, ``{Packet Drop
  and RSSI Evaluation for LoRa: An Indoor Application Perspective},'' in
  \emph{Proceedings of IEEE 7th World Forum on Internet of Things (WF-IoT)},
  2021, pp. 913--914.

\bibitem{Rando2021}
R.~Patgiri, ``{Rando: A General-purpose True Random Number Generator for
  Conventional Computers},'' in \emph{Proceedings of IEEE 20th International
  Conference on Trust, Security and Privacy in Computing and Communications
  (TrustCom)}, 2021, pp. 107--113.

\bibitem{jacak2021quantum}
M.~M. Jacak, P.~J{\'o}{\'z}wiak, J.~Niemczuk, and J.~E. Jacak, ``Quantum
  generators of random numbers,'' \emph{Scientific Reports}, vol.~11, no.~1,
  pp. 1--21, 2021.

\bibitem{kietzmann2021guideline}
P.~Kietzmann, T.~C. Schmidt, and M.~W{\"a}hlisch, ``A guideline on pseudorandom
  number generation (prng) in the iot,'' \emph{ACM Computing Surveys (CSUR)},
  vol.~54, no.~6, pp. 1--38, 2021.

\bibitem{PIM2021}
V.~P. Yanambaka, A.~Abdelgawad, and K.~Yelamarthi, ``Pim: A puf-based host
  tracking protocol for privacy aware contact tracing in crowded areas,''
  \emph{IEEE Consumer Electronics Magazine}, vol.~10, no.~4, pp. 90--98, 2021.

\bibitem{amsaad2021enhancing}
F.~Amsaad, A.~Oun, M.~Y. Niamat, A.~Razaque, S.~Kose, M.~Mahmoud, W.~Alasmary,
  and F.~Alsolami, ``Enhancing the performance of lightweight configurable puf
  for robust iot hardware-assisted security,'' \emph{IEEE Access}, vol.~9, pp.
  136\,792--136\,810, 2021.

\bibitem{rahman2017hardware}
F.~Rahman, M.~Farmani, M.~Tehranipoor, and Y.~Jin, ``{Hardware-assisted
  cybersecurity for iot devices},'' in \emph{Proceedings of 18th International
  Workshop on Microprocessor and SOC Test and Verification (MTV)}.\hskip 1em
  plus 0.5em minus 0.4em\relax IEEE, 2017, pp. 51--56.

\bibitem{mohanty2021introduction}
S.~P. Mohanty, J.~Plusquellic, G.~S. Rose, W.~Zhang, and M.~K. Michael,
  ``Introduction to the special issue on hardware-assisted security for
  emerging internet of things,'' pp. 1--3, 2021.

\bibitem{Yanambaka2016}
V.~P. Yanambaka, S.~P. Mohanty, E.~Kougianos, and J.~Singh, ``{Secure Multi-key
  Generation Using Ring Oscillator Based Physical Unclonable Function},'' in
  \emph{Proceedings of IEEE International Symposium on Nanoelectronic and
  Information Systems (iNIS)}, 2016, pp. 200--205.

\bibitem{chaoticRNG2020}
S.~Kalanadhabhatta, D.~Kumar, K.~K. Anumandla, S.~A. Reddy, and A.~Acharyya,
  ``Puf-based secure chaotic random number generator design methodology,''
  \emph{IEEE Transactions on Very Large Scale Integration (VLSI) Systems},
  vol.~28, no.~7, pp. 1740--1744, 2020.

\bibitem{idris2021}
T.~A. Idriss, H.~A. Idriss, and M.~A. Bayoumi, ``A lightweight puf-based
  authentication protocol using secret pattern recognition for constrained iot
  devices,'' \emph{IEEE Access}, vol.~9, pp. 80\,546--80\,558, 2021.

\bibitem{yuang2021}
P.~Yuan, B.~Li, Y.~Zhang, J.~Wu, H.~Zheng, and C.~Wang, ``{A PUF-Based
  Lightweight Broadcast Authentication Protocol for Multi-Server Systems Using
  Blockchain},'' in \emph{Proceedings of IEEE 6th International Conference on
  Signal and Image Processing (ICSIP)}, 2021, pp. 1035--1041.

\bibitem{mall2022}
P.~Mall, R.~Amin, A.~K. Das, M.~T. Leung, and K.-K.~R. Choo, ``Puf-based
  authentication and key agreement protocols for iot, wsns and smart grids: A
  comprehensive survey,'' \emph{IEEE Internet of Things Journal}, pp. 1--1,
  2022.

\bibitem{6823677}
C.~Herder, M.-D. Yu, F.~Koushanfar, and S.~Devadas, ``Physical unclonable
  functions and applications: A tutorial,'' \emph{Proceedings of the IEEE},
  vol. 102, no.~8, pp. 1126--1141, 2014.

\bibitem{9595111}
K.~Khalil, A.~Abdelgawad, and M.~Bayoumi, ``Intelligent resource discovery
  approach for the internet of things,'' in \emph{2021 IEEE 7th World Forum on
  Internet of Things (WF-IoT)}, 2021, pp. 264--269.

\bibitem{puf}
M.~I. Mahmud, P.~K. Sadhu, V.~P. Yanambaka, and A.~Abdelgawad, ``Vxorpuf: A
  vedic principles - based hybrid xor arbiter puf for robust security in
  iomt.'' Preprints.org 2023, 2023030499.
  https://doi.org/10.20944/preprints202303.0499.v1.

\end{thebibliography}

\end{document}